\documentclass[pre,reprint,twocolumns,notitlepage,superscriptaddress]{revtex4-1}
\usepackage{blindtext}
\usepackage{hyperref}
\usepackage{amsmath,ulem, bm}
\usepackage{lipsum}
\usepackage{eucal}
\usepackage[usenames,dvipsnames]{color}
\usepackage{graphicx}
\usepackage[toc,page]{appendix}
\usepackage{amssymb,amsfonts,amsmath}
\usepackage{bm}
\usepackage{enumerate}
\usepackage{graphicx}
\usepackage[caption=false]{subfig}
\usepackage[toc,page]{appendix}
\usepackage{comment}

\begin{document}
\author{A.~Sanzeni}
\affiliation{Department of Physics, University of Milan and INFN, Via Celoria 13, 20133 Milano, Italy}
\affiliation{Department of Physics, University of California San Diego, La Jolla, CA 92093-0374, USA}
\author{V.~Balasubramanian}
\affiliation{David Rittenhouse Laboratory, University of Pennsylvania,  Philadelphia, PA 19104, USA}
\author{G.~Tiana}
\affiliation{Centre for Complexity \& Biosystems and Department of Physics, University of Milan and INFN,  University of Milan, via Celoria 16, 20133 Milano, Italy}
\author{M.~Vergassola}
\affiliation{Department of Physics, University of California San Diego, La Jolla, CA 92093-0374, USA}

\title{Complete coverage of space favors modularity  \\ 
of the grid system in the brain}

\begin{abstract}
Grid cells in the entorhinal cortex fire when animals  that are exploring a certain region of space occupy the vertices of a triangular grid that spans the environment.  
Different neurons feature triangular grids that differ in their properties
of periodicity, orientation and ellipticity. Taken together, these grids allow the animal to maintain an internal, mental representation of physical space.
Experiments show that grid cells are modular, i.e. there are groups of neurons 
which have grids with similar periodicity, orientation and ellipticity. 
We use statistical physics methods to derive a relation between variability of the properties of the grids within a module and the range of space that can be covered completely (i.e. without gaps) by the grid system with high probability.   
Larger variability shrinks the range of representation, providing a functional rationale for the experimentally observed co-modularity of grid cell periodicity, orientation and ellipticity.  
We obtain a scaling relation between the number of
neurons and the period of a module, given the variability and coverage range. 
Specifically, we predict how many more neurons are required at smaller grid scales than at larger ones. 
\end{abstract}
\maketitle

\section{Introduction} 
Classical behavioral experiments show that the navigation of mammals relies on an internal representation of space called a ``cognitive map''~\cite{tolman1948}. 
Research on the neural basis of this internal representation started with the discovery of place cells in the hippocampus of rats, neurons that have their activity controlled by the physical position occupied by the animal~\cite{Okeefe}. The discovery of place cells generated an extensive investigation  of the spatial representation system in the brain, which led to the discovery of grid cells \cite{Hafting2005}, as well as
different types of neurons whose activity codes for head direction~\cite{Ranck84}, speed~\cite{Kropff:2015aa}, and borders of the environment~\cite{HIPO20511,Solstad1865} (see~\cite{Moser2014} for a recent review). 
The discovery of cells that constitute a positioning system in the brain was the motivation for the Nobel Prize in Physiology awarded in 2014.

\smallskip
One of the most striking elements composing the cognitive map is in the entorhinal cortex (EC) \cite{Hafting2005}, where grid cells respond when the animal occupies one of the vertices of a triangular grid that tessellates space. 
It is widely believed that these neurons provide a metric for the spatial representation system, since their relation with physical position does  not reshuffle in different environments, unlike what happens for place cells where ``remapping''' occurs~\cite{Muller19871951}.

Grid cells are organized in modules -- grids in a module are clustered around a discrete period which increases along the dorso-ventral axis of the EC \cite{Barry2007,Stensola2012}. 
Grids in a module also share similar orientations and ellipticities while varying in spatial phase \cite{Stensola2012}.  
Experimentally, there is a  power-law relation between the periodicities of different modules -- a rationale for the power law was given in  \cite{Wei2013,mathis2012}. 

The term ``module'' for grid cells differs fundamentally from the same term used in the context of  brain (or city) networks \cite{Barthelmy}. There, neurons correspond to the nodes of a network whose edges correspond to axonal connections among the neurons. Modularity refers to the formation of clusters of nodes that are more densely connected among themselves than to nodes in other modules. The reason for modularity in these networks is that edges have costs that scale with their length, so that spatial aspects are important and commonly lead to cluster formation.   By contrast, the triangular lattices in the EC grid system describe firing patterns of  {\it individual} grid cell neurons as an animal explores the environment. In other words, there is no physical edge between vertices of the triangular firing lattices of grid cells, and no cost associated to their length. Hence, physical proximity is not relevant for grid cells and has no bearing on the problem of explaining the modular organization of grid cells in the EC.

Why is the grid system modular? The key point underlying our arguments in the sequel is that behavioral deficits in orientation and navigation  result if the neural representation has gaps, i.e. complete coverage of space is a fundamental requirement for the cognitive map to function.
Specifically, we use statistical physics methods to show that variability in period, orientation or ellipticity randomizes the relative phases of firing fields, and leads to failure of spatial coverage. 
Larger variability entails a smaller physical range that can be covered without gaps (which would lead to behavioral deficits). 
Hence, optimizing spatial coverage gives a functional argument for reduced variability and for the observed 
co-modularity of grid cells. 
We also predict a scaling law relating the period and number of neurons in a module.

\section{Results}
\subsection{A model of grid cells' activity}
\begin{figure}[h]
  \centering
\includegraphics[width=8.5cm,keepaspectratio]{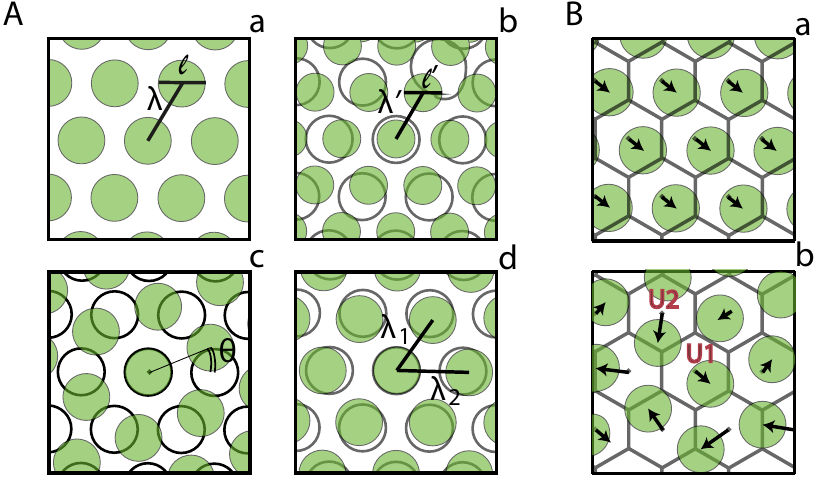}
    \caption{\small{\textbf{Deformations in grid parameters induce dephasing.} 
(\textbf{A}) Spatial activity of a grid cell  (\textbf{a}) and its possible deformations: dilation (\textbf{b}), rotation (\textbf{c}) and 
ellipticity transformation (\textbf{d}). The activity after transformation (green) is superimposed on the reference activity (gray).
 (\textbf{B}) Using a set of grid parameters we divide space into unit cells represented by gray hexagons.  A neuron with the 
 same set of  grid parameters (\textbf{a}) has constant relative phase (black arrows) in each unit cell.
A neuron with different grid parameters (\textbf{b}) has variable phases in different unit cells, e.g. the grid is rotated as in panel 
(\textbf{A.c}) and the center of the unit cell is covered by a firing field in U1 but not in U2.
  }}
\label{fig:Figure_1}
\end{figure}

For our specific purpose of analyzing efficient coverage of space, the firing field of grid cells can be simplified as follows. 
After thresholding for noise, the smooth lumps formed by firing fields are treated as being uniformly active inside a localized region and inactive outside (Fig.~\ref{fig:Figure_1}).  
Noise and firing inhomogeneity inside the active region only degrade the uniformity of coverage relative to this model.  Thus, treating firing fields as step functions allows us to derive bounds on how well a given grid architecture can cover space. 

Specifically, we represent the activity of grid cells as
\begin{equation}
\label{eq:activity}
a({\bf x})=\sum_{n, m \in \mathbb{Z}} \chi \left(\frac{|\boldsymbol{\phi}+R(\theta)\left[n  {\bf v}+m  {\bf u}\right)]- {\bf x}|}{\ell/2}\right)\, , 
\end{equation}
where ${\bf x}$ is the vector locating the position of the animal in two dimensions, ${\bf v}=\lambda_1(\cos(\beta),\sin(\beta))$ and ${\bf u}=\lambda_2(1, 0)$ are the elementary vectors that generate the grid, $\ell$ is the diameter of a firing field, and $n$ and $m$ are integers indexing the vertices of the grid.   $R(\theta)$ is an overall rotation of the grid by an angle $\theta$, the angle $\beta$ describes the relative rotation of the grid basis vector ${\bf v}$ relative to ${\bf u}$, and the phase $\boldsymbol{\phi}$ represents a shift with respect to a reference point.
The activity of an equilateral, unrotated triangular grid has  $\lambda_1=\lambda_2=\lambda$, $\beta=\pi/3$ and $\theta=0$. 
The set of the six vertices defined by the triplet ${\bf u}$, ${\bf v}$, ${\bf u}-{\bf v}$ and their opposite vectors forms an hexagon that can be inscribed into an ellipse. The ratio between the axes of the ellipse defines the ellipticity $\epsilon$ of the grid ($\epsilon=1$ for equilateral grids).  
Hereafter, we study isosceles grids, where the relation  $\cos \beta=1/\sqrt{1+3\epsilon^2}$ holds,  but our conclusions hold generally (see Appendix~\ref{supp:Ellipticity}).
Finally, for the purpose of analyzing coverage we take $\chi=1$ when its argument is $<1$ and $\chi=0$ otherwise, i.e. we are interested in whether a neuron is active or not at a given point (disregarding its strength of activity).

In a module, grid cells with similar spacing  have similar orientation, ellipticity and firing field size \cite{Stensola2012}. 
However, parameters of the grids have an appreciable variability, which we quantify using experimental
data reported in Stensola et al.~\cite{Stensola2012} as follows. 
For each animal where the distribution of grid parameters is available, we fit the data with a sum of Gaussians.
For each module, we used one Gaussian for the period (mean $\lambda$, standard deviation $\sigma_{\lambda}$) and one for the orientation (mean $\theta$, standard deviation $\sigma_{\theta}$.)
The two standard deviations are roughly constant in the various modules.
Indeed, the Pearson correlation coefficient is $0.21$ between $\sigma_{\lambda}$ and $\lambda$ and  $0.28$ between $\sigma_{\theta}$ and $\lambda$.
The standard deviation of the grid period is about $6$ cm. 
Assuming 10 modules and that the smallest is about $40$cm, the ratio $\sigma_{\lambda}/\lambda$ goes from $0.01$ to $0.15$.
The standard deviation of the orientation in a module is about $0.03$ rad. 
In the literature we were not able to find the distribution of ellipticity in the population within a single module. 
We know that  ellipticity also has a modular structure and that across a population (all modules) the mean ellipticity is around $1.16\pm0.003$  \cite{Stensola2012}.
In the following analysis we assume a standard deviation in ellipticity in the range $0.01$-$0.15$, i.e. similar to variability in grid spacing.
Finally, we fixed the ratio between the firing field width and the grid
spacing at the experimental value $\lambda/\ell\sim1.63\pm0.04$~\cite{Giocomo2011}.

\subsection{Dephasing and decorrelation of neuronal activity}\label{sec:dephasing}
In order to cover an environment with grid cells, there must be at least one active neuron at each point.  
The average orientation $\theta$, ellipticity $\epsilon$ and period $\lambda$ within a module define a tessellation of the plane into periodic unit cells. 
Perfect periodicity would imply that once a unit cell is covered, all of space is covered.  
However, perfect periodicity is broken by the variability discussed above, which results in deviations from the average grid. 
Indeed, as shown in Fig.~\ref{fig:Figure_1}B, the pattern of firing fields changes across unit cells, a phenomenon that we call  ``dephasing''.
Here we characterize this effect by computing the correlation coefficient between the number of neurons that are active at the center of two unit cells.

\begin{figure}[h]
    \centering
    \includegraphics[width=8.5cm,keepaspectratio]{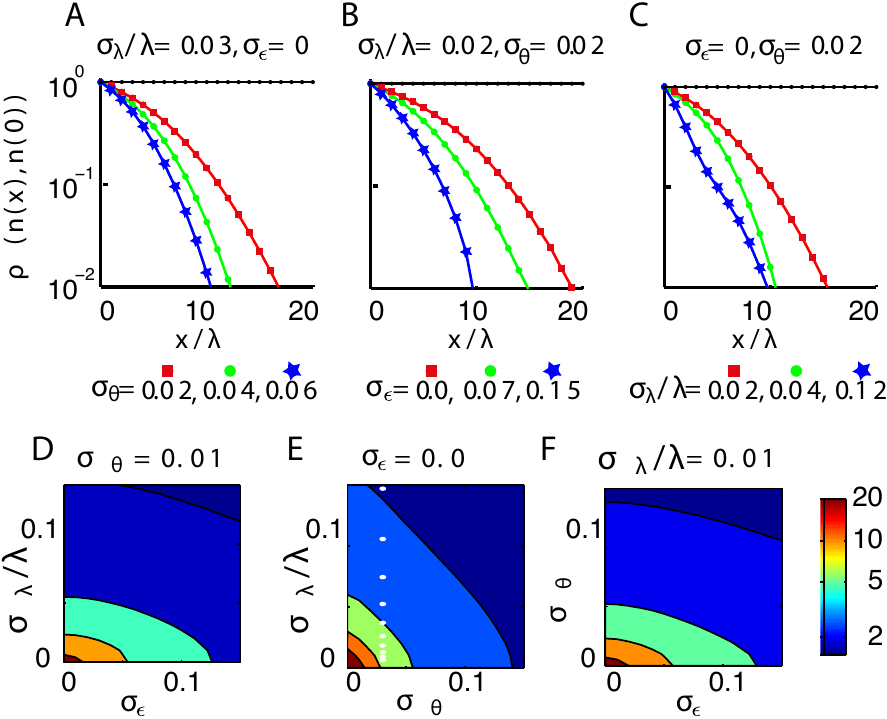}
\caption{\small{\textbf{Variability of grid parameters induces decorrelation in grid cells activity.} 
The correlation coefficient of the number of neurons active at the center of two unit cells along the line ${\bf x}=n{\lambda}(1,0)$, $n\in \{0,\, 1,\, \dots\}$ for different values of the standard deviations in the parameters of the grid cells.  
In the plots $x = |{\bf x}|$.
Colors in the first row represent different values of one of the variances (orientation (\textbf{A}), ellipticity (\textbf{B}), period (\textbf{C})) while the other two are fixed ($\sigma_{\lambda}/\lambda=0.03$, $\sigma_{\epsilon}=0$(\textbf{A}), $\sigma_{\lambda}/\lambda=0.02$, $\sigma_{\theta}=0.02$ (\textbf{B}), $\sigma_{\theta}=0.02$, $\sigma_{\epsilon}=0$ (\textbf{C})). The case with no noise is represented in black.
Larger variances lead to a rapid decrease in the correlation as a function of separation.   (See Appendix~\ref{supp:correlation_function} for details on panel \textbf{C}.)
The correlation length depends on three variances which we varied in pairs obtaining contour plots ($\sigma_{\theta}=0.01$ (\textbf{D}), $\sigma_{\epsilon}=0$ (\textbf{E}), $\sigma_{\lambda}/\lambda=0.01$ (\textbf{F})).
The white points in panel (\textbf{E}) correspond to the values of the standard deviations  measured in~\cite{Stensola2012}.
 }}
\label{fig:Figure_2}
\end{figure}

\smallskip
The number of neurons active at a spatial point ${\bf x}$ is given by $n({{\bf x}})=\sum_{i=1}^N a_i({\bf x})$, where $a_i$ is the spatial activity of the $i-$th neuron given by Eq.~\eqref{eq:activity} and $N$ is the number of neurons in the system.   
Consider a set of neurons whose grid parameters are drawn from Gaussian distributions with  standard deviations $\sigma_{\lambda}$, $\sigma_{\theta}$ and $\sigma_{\epsilon}$.  
We compute numerically the correlation  (normalized to $1$ for coincident points) between the numbers of neurons  $n({\bf x})$ and $n({\bf y})$ active at different points ${\bf x}$ and ${\bf y}$, by averaging over statistical realizations (Fig.~\ref{fig:Figure_2}).   
The correlation declines systematically with the separation in the grid lattice. 
The corresponding correlation length $\mathcal{L}$ (defined as the distance at which the correlation drops to  $1/e$)  
decreases with the variance in the parameters of the grid cells (Fig.~\ref{fig:Figure_2}) and  is in the meter scale for the smallest modules, which is within the behavioral range of a few tens of meters found in rats \cite{davis1948studies,braun1985home,slade1983home}.

\smallskip

We can understand the asymptotic behavior of the correlation function of the number of neurons active at two spatial points at large separations as follows.  We are assuming that grid cells in a module fire independently.  Therefore, 
 the correlation function of the number of neurons active at ${\bf x}$ and ${\bf y}$,  $\rho({n(\bf{x}}),n({\bf{y}}))$, is equal to the correlation function of the activity of a single generic neuron $a({\bf x})$, averaged over the distribution of grid parameters, $\rho({a(\bf{x}}),a({\bf{y}}))$. 
The mean activity of a single neuron is obtained from Eq.~\eqref{eq:activity} by averaging over the grid parameters.
This quantity can be written as 
\begin{equation}
\langle a({\bf x})\rangle =  \int  \,  a({\bf x}) \, d P_{\theta}d P_{\lambda} d P_{\epsilon}  d P_{\boldsymbol{\phi}} \, , 
\end{equation}
where $d P_{(*)}$ represents the probability distribution for the parameter ${(*)}$.
As discussed above, orientation, period and ellipticity of the grids follow a Gaussian distribution whilst the spatial phase is uniformly distributed in a unit cell.
To compute the integral, we divide space into unit cells and consider the center of the cell containing ${\bf x}$ as a reference point for  the phase of the grid $\boldsymbol{\phi}$.
Because $\boldsymbol{\phi}$ is uniformly distributed in the unit cell, and because $\chi=1$ within the firing field and $\chi=0$ outside, the integral over  $\boldsymbol{\phi}$ is a constant equal to the ratio between the area of a firing field and that of a unit cell, i.e. $\pi/2\sqrt{3} \, \left(\ell/\lambda\right)^2$.
The remaining  integrals are equal to unity and we finally obtain $\langle a({\bf x})\rangle=\pi/2\sqrt{3} \, \left(\ell/\lambda\right)^2$.

In order to compute the correlation function we need to determine the quantity $\langle a({\bf{y}})a({\bf{x}})\rangle$.
which can be written as 
\begin{eqnarray}
\langle a({\bf{y}})a({\bf{x}})\rangle= \int    Q_{\boldsymbol{\phi}}({\bf{y}},{\bf{x}}) d P_{\boldsymbol{\phi}} \,;\nonumber \\
 Q_{\boldsymbol{\phi}}({\bf{y}},{\bf{x}}) \equiv \int   a({\bf{y}})a({\bf{x}}) \, d P_{\theta}d P_{\lambda} d P_{\epsilon}\,  ,
\end{eqnarray}
where $ Q_{\boldsymbol{\phi}}({\bf{y}},{\bf{x}})$ is the joint probability distribution that a neuron is active both at ${\bf{y}}$ and at ${\bf{x}}$. The distribution depends parametrically on  ${\boldsymbol{\phi}}$. 
The joint probability can be computed as
\begin{equation}
Q_{\boldsymbol{\phi}}({\bf{y}},{\bf{x}})=Q_{\boldsymbol{\phi}}({\bf{y}}|{\bf{x}}) Q_{\boldsymbol{\phi}}({\bf{x}})\,,
\end{equation}
where 
$Q_{\boldsymbol{\phi}}({\bf y}|{\bf{x}})$ 
is the conditional probability that a neuron is active at ${\bf{y}}$  if it is active at ${\bf{x}}$ (for a given ${\boldsymbol{\phi}}$). 
The quantity  $Q_{\boldsymbol{\phi}}({\bf x})$ is the probability that a neuron is active at  ${\bf{x}}$, again for a given ${\boldsymbol{\phi}}$.

In order to evaluate the conditional probability $Q_{\boldsymbol{\phi}}({\bf{y}}|{\bf{x}})$, we divide space into unit cells using the mean grid parameters.
We consider a neuron with ${\boldsymbol{\phi}}=(0,0)$, i.e. with a firing field centered at the origin, and analyze the evolution of its phase ${\boldsymbol{\phi}}_n$ in the unit cells centered at ${\bf y}=(n\lambda,0)$, $n=0, \, 1, \, \dots$.
If the grid cell has the same grid properties as the average grid, its phase will be invariant, i.e. ${\boldsymbol{\phi}}_n={\boldsymbol{\phi}}$, hence  $Q_{\boldsymbol{\phi}}({\bf y}|{\bf{x}})=1$ and the correlation function will be a constant that does not depend ${\bf y}$.
If there is variability, the phase of the grid will be randomly distributed in the two-dimensional area of the unit cell centered at $(n\lambda,0)$ as $n$ increases. 
It follows that $Q_{\boldsymbol{\phi}}({\bf y}|{\bf{x}})\to \pi/2\sqrt{3} \, \left(\ell/\lambda\right)^2$;  $\langle a({\bf{y}})a({\bf{x}})\rangle \to\langle a({\bf{x}})\rangle^2$ and the correlation asymptotically goes to zero as shown in Fig.~\ref{fig:Figure_2}.

In Appendix~\ref{supp:correlation_function} we discuss the behavior of the correlation function in the absence of orientation variance.
This analysis is not relevant to describe the biological system, where orientation variance is estimated to be about 0.03 rad, but constrains the definition of the correlation length.
In particular, we show that the threshold used to define the correlation length ought to be in the range $[0.28,1]$, which includes our choice of a threshold equal to $1/e$.

\subsection{Coverage drives modularity}
In order to cover an environment with a set of grid cells, there has to be at least one active neuron at every point.
The correlation length $\mathcal{L}$ characterizes the scale beyond which the numbers of active neurons become approximately independent. 
A region of size $R$ is thereby decomposed in  $R^2/\mathcal{L}^2$  regions whose
coverage probabilities are roughly independent of each other. 
If each of these is covered with probability $p$, the  probability $P$ of covering the whole environment is 
\begin{equation}\label{eq:p_of_R_of_L}
\log (P)=\gamma { \frac {R^2}{\mathcal{L}^2}} \log(p)\, ,
\end{equation}
where $\gamma$ is a constant that depends on the geometry of the system.

\begin{figure}[htb]
  \centering
 \includegraphics[keepaspectratio, keepaspectratio]{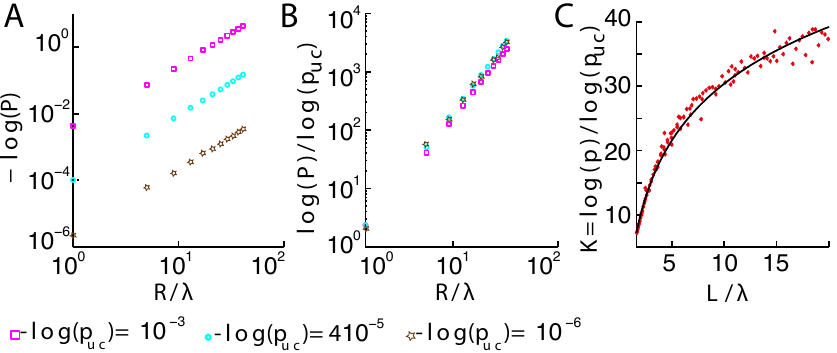}
  \caption{\textbf{Covering probability of a correlation volume of linear size $\mathcal{L}$.}
  (\textbf{A}) Covering probability $P$ of a circular environment of linear size $R$ computed numerically for different environment size and covering probability of the unit cell $p_{uc}$.
 The different values of $p_{uc}$ have been obtained using sets of grid cells made of a different number of neurons.
  (\textbf{B}) Results of panel (\textbf{A}) collapse on a common curve when their logarithm is rescaled by $\log(p_{uc})$, justifying the functional form introduced in Eq.~\ref{eq:functional_form}.
  (\textbf{C}) Numerical covering probability $p$ of a correlation volume of linear size $\mathcal{L}$  have been used to obtain an empirical description of  the function  $\mathcal{K}\left(\mathcal{L}/\lambda \right)$ described in Eq.~\ref{eq:functional_form} (red dots).
  The best fit (black line) is given by the function $\mathcal{K}\left(x \right)= 0.73 +13 \log(x)$.
Simulations parameters are  $\lambda/\ell=$1.63, $\sigma_{\theta}=0.04$, $\sigma_{\epsilon}=0$.
In panels (\textbf{A}-\textbf{B}) we fixed $\sigma_{\lambda}/\lambda=0.08$   whilst the number of neurons $N$ is 30 (magenta, squares), 40 (light blue, circles), 50 (brown, stars). 
In panel (\textbf{C}) we fixed  $N=30$ and $\sigma_{\lambda}/\lambda$ has been varied to span the different values of $\mathcal{L}$ observed in the biological system as described in Fig~\ref{fig:Figure_2}E.
}
  \label{fig:Figure_3}
\end{figure}

\smallskip
The probability $p$ of covering a correlation volume of linear size $\mathcal{L}$ as a function  of the probability of covering a unit cell $p_{uc}$, was  obtained numerically as follows.
We computed  the covering probability of a circular environment of radius $R$ using sets of grid cells characterized by different $p_{uc}$. 
Results of the simulations are shown in Fig.~\ref{fig:Figure_3}.
For every value of  the radius $R$ the logarithm of the covering probability rescaled over $\log(p_{uc})$ does not depend on $p_{uc}$  (Fig.~\ref{fig:Figure_3}B).
It follows that the probability $p$ of covering a correlation volume of linear size $\mathcal{L}$ can be expressed as 
\begin{equation}\label{eq:functional_form}
\log(p)=\mathcal{K} \, \log(p_{uc}) \, ,
\end{equation}
where $\mathcal{K}$ is a function that depends only on $\mathcal{L}/\lambda$ for dimensional reasons.

Over a range of grid variances that includes the experimentally measured values, we found that $\mathcal{L}/\lambda \lesssim 20$ (Fig.~\ref{fig:Figure_2}).
In this range, we  found numerically that 
\begin{equation}\label{eq:kappa}
\mathcal{K}\left(x \right)= c_1 +c_2 \log(x)
\end{equation}
($c_1=0.73$, $c_2=13$) gives a good description of the data (see Fig.~\ref{fig:Figure_3}C) .

\begin{figure}[h]
    \centering
    \includegraphics[width=8.5cm,keepaspectratio]{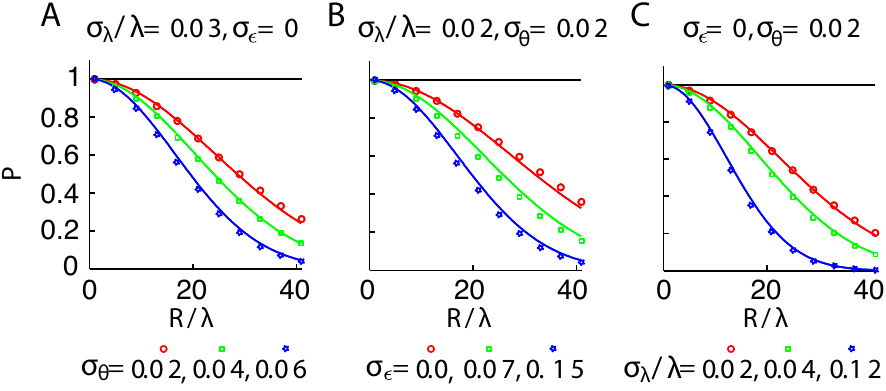}
    \caption{\small{\textbf{The covering probability $P$ decreases with the variance of grid parameters and the size of the environment.}
    $P$ is computed for circular environments of radius $R$ and for different grid variances.     
    Colors represent different values of one of the variances (orientation (\textbf{A}), ellipticity (\textbf{B}), period (\textbf{C})) while the other two are fixed  ($\sigma_{\lambda}/\lambda=0.03$, $\sigma_{\epsilon}=0$(\textbf{A}), $\sigma_{\lambda}/\lambda=0.02$, $\sigma_{\theta}=0.02$ (\textbf{B}), $\sigma_{\theta}=0.02$, $\sigma_{\epsilon}=0$ (\textbf{C})). The case with no noise is  in black.
Numerical results  (colored symbols) 
match theoretical predictions (continuous line) obtained by Eq.~(\ref{eq:p_of_R}).
The number of neurons  $N=30$.
     }}
\label{fig:Figure_4}
\end{figure}

Combining  Eqs.~\eqref{eq:p_of_R_of_L}, \eqref{eq:functional_form} and~\eqref{eq:kappa}, we obtain\,:
\begin{equation}\label{eq:p_of_R}
\log (P)=\gamma { \frac {R^2}{\mathcal{L}^2} \mathcal{K} \left(\frac{\mathcal{L}}{\lambda} \right)}\, \log(p_{uc})\,. 
\end{equation}
To test this estimate, we numerically analyzed the covering probability of a circular environment of radius $R$ by $N$  neurons whose grid parameters are drawn from Gaussian distributions.
We then checked if every point in the environment is covered by at least one grid cell and we averaged over realizations.  
Fig.~\ref{fig:Figure_4} confirms the validity of Eq.~(\ref{eq:p_of_R}),
with a proportionality constant $\gamma=0.804$.

 \smallskip
 Fig.~\ref{fig:Figure_4} and Eq.~(\ref{eq:p_of_R})  show that the covering probability of a region increases with the correlation length.   In this sense,  a set of grid cells with a larger correlation length is more efficient, because with the same number of neurons, and hence a fixed $p_{uc}$, it will have fewer gaps.  Since the correlation length decreases if  the  standard deviations increase, we conclude that coverage drives modularity --  grid cells with similar period should have similar orientations and ellipticities as observed experimentally \cite{Stensola2012}. 
 
 \subsection{Gaps decline exponentially with the number of neurons}
We now quantify how the number of neurons $N$ in a module affects the probability of covering a range $R$. The dependence on $N$ in Eq.~\eqref{eq:p_of_R} occurs through the
factor $p_{uc}$. 
The random distribution of phases of grid cells \cite{Hafting2005} dictates an exponential dependence between the probability $p_{uc}$  and the number of neurons $N$.
Indeed, consider $N$ neurons that cover a unit cell of a $d$-dimensional grid with a single gap. 
An additional neuron added with a random phase will fail to overlap the gap with some probability $h<1$. 
If we add $Q$ additional neurons independently, the probability that they all miss the
gap is $h^Q$, i.e. the probability of gap persistence declines exponentially with the number of added neurons. 
Subleading terms  are captured by analyzing partial coverage with each additional neuron (see Appendix~\ref{supp:p_1D}).

\begin{figure}[h]
   \centering
  \includegraphics[width = 8.5cm,keepaspectratio]{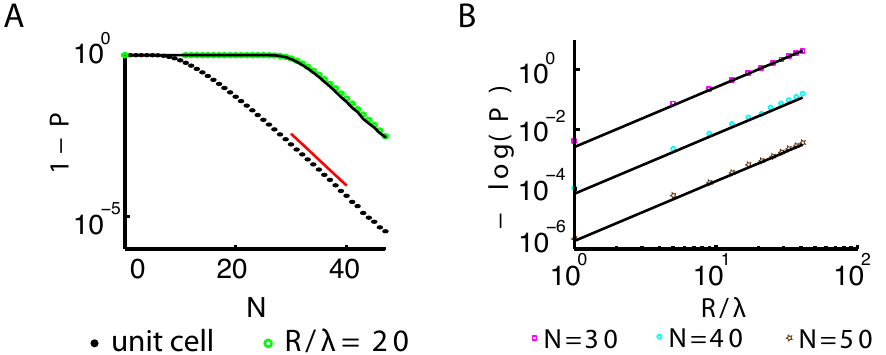}
    \caption{\small{
    \textbf{The covering probability $P$ increases with the number of neurons $N$.}
(\textbf{A}) The numerical computation of $P$  for a unit cell (black dots) 
is combined with Eq.~(\ref{eq:covering_probability}) to predict $P$ for an environment of size $R/\lambda=20$ (black line). Results of numerical simulations  are in green.
The function $1-P$ asymptotically decays as $\exp(-\alpha N)$ with $\alpha \approx 0.4$ (red line).
(\textbf{B}) The probability $P$ {\it vs.} the
environment size $R$ for different $N$. Results of the simulations (colored symbols)
match predictions (black lines)  by Eq.~\eqref{eq:covering_probability}.
Parameters are $\lambda/\ell=$1.63, $\sigma_{\theta}=0.04$, $\sigma_{\lambda}/\lambda=0.08$, $\sigma_{\epsilon}=0$.
    }}
\label{fig:Figure_5}
\end{figure}

Thus, for a large number of neurons in a two-dimensional grid module, we expect that $p_{uc}\sim1-\exp\left(-\alpha N\right)$, where $\alpha$ is a positive constant that depends, by dimensional analysis, on the ratio $\ell/\lambda$.
In the opposite limit, when $N$ is smaller than the area of the unit cell divided by the area of the firing field,  coverage cannot be achieved and $p_{uc}=0$, as confirmed numerically in Fig.~\ref{fig:Figure_5}.

In summary, the estimate for the probability $P$ of covering a two dimensional circular region of radius $R$ is 
\begin{equation}\label{eq:covering_probability}
\log (P)=\gamma  \frac {R^2}{\mathcal{L}^2} \mathcal{K} \left( \mathcal{L}/\lambda \right)\, \log(1-e^{\mathcal{F}\left(N\right)}) \, ,
\end{equation}
where the function $\mathcal{F}\left(N\right)$ behave as just discussed, which is 
 validated by numerical simulations (Figs.~\ref{fig:Figure_4}, \ref{fig:Figure_5}).  
On the one hand, the probability of gaps in coverage declines exponentially with $N$. 
On the other hand, the probability of gaps in coverage of a range $R$ increases exponentially as $(R/{\cal L})^2\mathcal{K} \left( \mathcal{L}/\lambda \right)$, where ${\cal L}$ decreases as the variability in a module increases.    
Hereafter, we balance these two effects to estimate the number of neurons required to cover space in modules of different mean periods.

\subsection{Prediction: smaller period modules need more neurons}
Eq.~\eqref{eq:covering_probability} gives the relation between the number of neurons $N_i$ and the parameters of the $i$-th module. Since the different modules vary systematically in their period, this relation predicts an associated variation in the number of neurons.

\smallskip
Assume that an animal encodes position within a region of size $R^2$ that is common to all the modules, and that the probability of covering space is the same at all scales.   
As we showed above, the probability of gaps in coverage declines exponentially with the number of neurons, 
and the coefficient in the exponent depends on the ratio $\ell / \lambda$ between the grid field 
width and the period. 
It is established experimentally that this ratio is fixed among modules~\cite{Barry2007,Stensola2012}.
Thus we can evaluate the predicted fraction of neurons in a given module, $N_i/\sum_i N_i$, where the denominator is a sum over modules, and $N_i$ is obtained by inverting Eq.~\eqref{eq:covering_probability}. 

\begin{figure}[h]
  \centering
  \includegraphics[width=8.5cm,keepaspectratio]{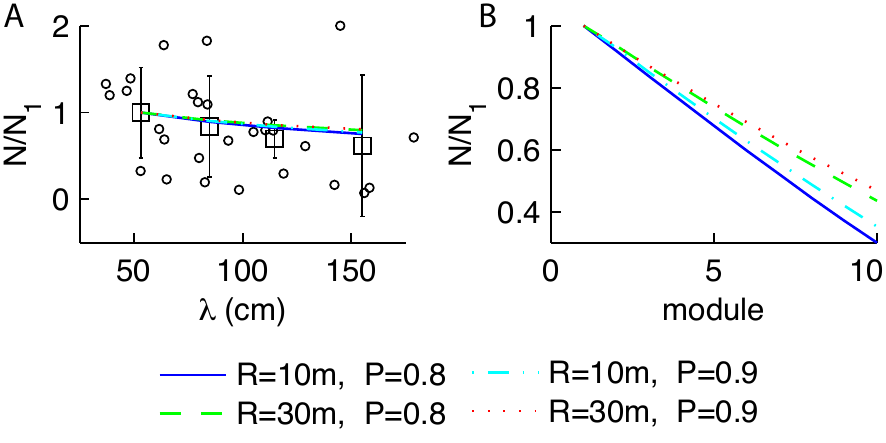}
    \caption{\small{\textbf{Number of neurons required for coverage decreases with the spatial period.} 
    (\textbf{A})  
We used experimental data in \cite{Stensola2012} to estimate the number of recorded neurons {\it vs} their spatial period (circles).
We applied k--means clustering to identify four modules according to the spatial period. 
For each module we computed the associated number of neurons and plotted the mean and the standard deviation of the number of neurons $N$ normalized over the number $N_1$ of neurons for the first modulus (black squares and lines).
Theoretical predictions given by Eq.~(\ref{eq:covering_probability})  obtained  with  different values of $R$ and  $P$ (lines) are compatible with  experimental data.
 (\textbf{B}) We extrapolated the number of neurons over ten modules for different values of $R$, $P$. 
Simulation parameters are $\lambda/\ell=1.63$, $\sigma_{\theta}=0.04$, $\sigma_{\epsilon}=0$ and  $\sigma_{\lambda}=6$cm.
}}
\label{fig:Figure_6}
\end{figure}

The results of this prediction and a comparison with the extant experimental data are
shown in Fig.~\ref{fig:Figure_6}. The theoretical predictions are given for a variety of ranges and
coverage probabilities, with the grid periods and variabilities fixed from experimental data.
Qualitatively, the theory predicts for any choice of parameters that the number of neurons
should decline with the period of the module, as also suggested by the data.

Responses from 4--5 modules spanning up to $50\%$ of the dorsoventral extent of mEC feature a smallest period of about $40$cm and a ratio of $1.42$ between consecutive scales \cite{Stensola2012}. 
This suggests that there should be about 10 modules in total in the rat grid system with a maximum period of about $10$m.
Fitting our theoretical predictions to experimental data \cite{Stensola2012}, we found that a range of a few tens of meters can indeed be covered with a high coverage probability in the range 80-90$\%$.  
Within the range of parameters that allows this coverage in our model, we predict a decrease of about {$50-70\%$} in the number of neurons between the first and the tenth module (Fig.~\ref{fig:Figure_6}).  
Experimental uncertainties and possible biases in recording from harder-to-reach modules with larger periods, prevent 
stringent fits.  Nevertheless, our theory robustly states that the number of neurons should decline with the period of the grid module.   

\section{Discussion}
A striking experimental observation about the grid system in the entorhinal cortex is that it is organized in discrete modules that share similar periods, orientations and ellipticities~ \cite{Stensola2012}.
Given this modular structure, the geometric progression of grid periods can be shown to minimize the number of neurons required to provide a specified spatial resolution~\cite{mathis2012,Wei2013}.
However, why would a modular architecture be necessary in the first place? In this paper, we have shown that efficient coverage of space favors modularity.

To study how variability in the grid parameters within a module would affect the probability of holes in coverage, we simply asked whether each
neuron did or did not fire above threshold at a given location. 
Alternatively, we could sum firing profiles of grid cells to assess how grid variability affects homogeneity of the population firing across space. 
Again, the key variable would be the correlation in the expected number of action potentials at each point in space.
The overall probability of coverage would be determined by a product of factors over each correlation volume, leading to the same conclusion.

We chose to analyze coverage because any grid coding scheme, e.g.,  \cite{Fiete2008,Burak2009,Sreenivasan2011,mathis2012,andreasetal,Wei2013,bush2015using}, requires neurons to be active at each point in space.  
Thus, we view our approach as setting a minimal requirement for a functioning grid system for encoding location.   
Our model predicted that there would be fewer neurons in modules with larger periods. We compared our theory with the actual numbers of neurons recorded across modules, which  
should be taken with caution because of potential biases in the recording methods, especially for deeper structures in the brains.  
Some additional evidence for a decrease of neurons with the period of modules stems from the relatively smaller size of the ventral entorhinal cortex (which is enriched in large periods) relative to the dorsal region.   
Indirect evidence also comes from the larger drifts seen in the activity of grid cells with larger periods \cite{hardcastle}\,:  attractor models indeed predict that networks with smaller numbers of neurons will drift more.  Further data is needed to confirm 
these indirect lines of evidence.  
Comprehensive recordings from many grid modules are challenging because modules of a given period are  not strictly localized anatomically, and because ventral regions are harder to record from. But such data will greatly illuminate models of the functional logic of the grid system, and will further test our quantitative predictions.

\begin{acknowledgments}
VB was supported by the Fondation P.G. de Gennes at the ENS, Paris;  by NSF grant  PHY-1066293 at the Aspen Center for Physics; and by NSF PoLS grant PHY-1058202. VB and MV were supported by NSF Grant PHY11-25915 at the KITP, Santa Barbara.
 \end{acknowledgments}

\appendix


\makeatletter 
\renewcommand{\thefigure}{S\@arabic\c@figure}
\makeatother
\setcounter{figure}{0}
\section{Covering probability of non-isosceles grids}\label{supp:Ellipticity}
In the main text we analyzed the covering probability of the grid system assuming isosceles grids; in this Section we show that our results hold also in the case of general grids.

\begin{figure}[htb]
  \centering
   \includegraphics[keepaspectratio,keepaspectratio]{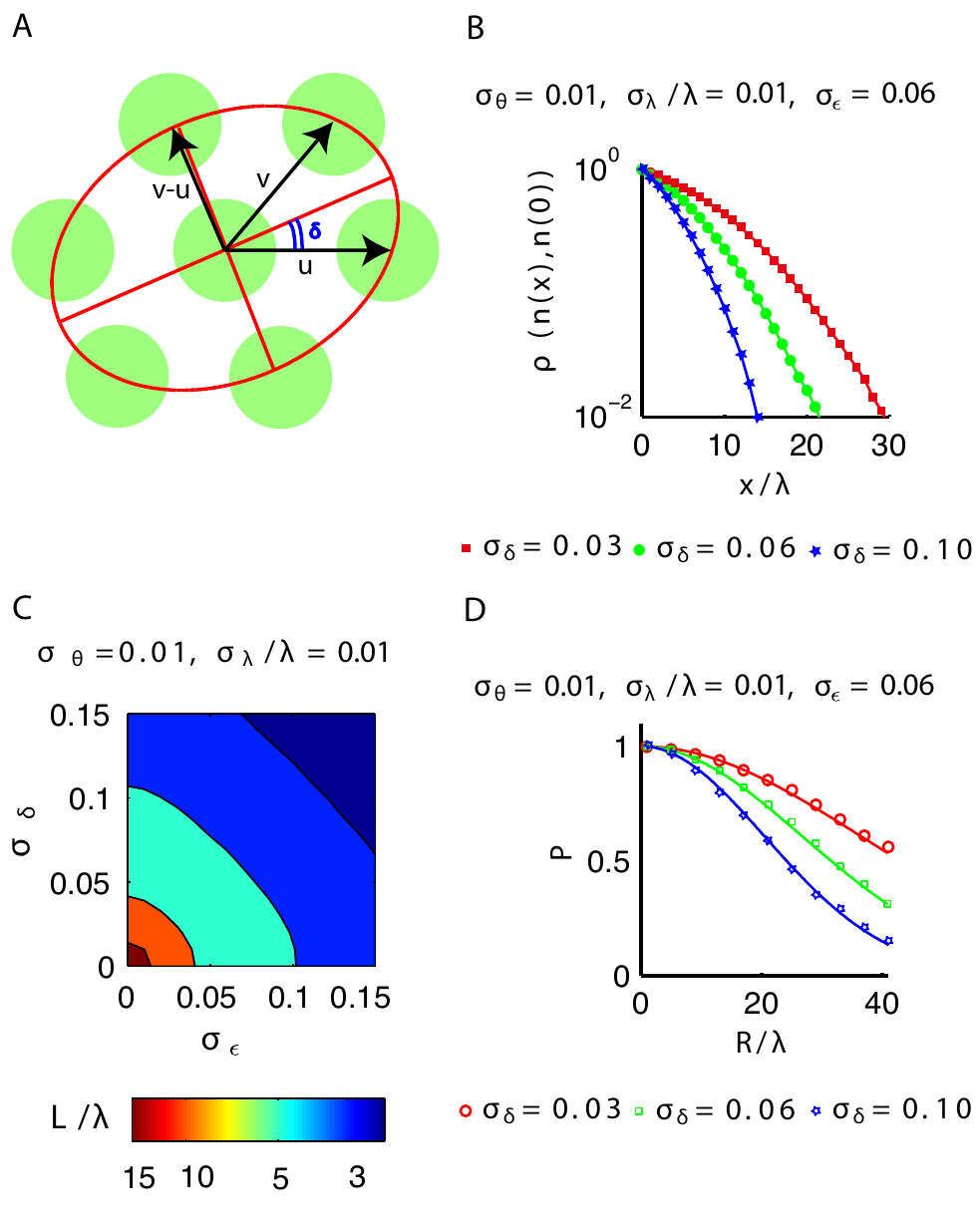}
\caption{\textbf{Variability in ellipticity $\epsilon$ and  direction $\delta$ induces decorrelation of activity and decreases the covering probability.}
We perform the same analysis as in the main text for the more general case $\delta\not=0$. 
({\bf{A}}) Representation of the grid activity with $\delta\not=0$.
The six firing fields defined by the triplet ${\bf u}$, ${\bf v}$, ${\bf u}-{\bf v}$ and their opposite vectors (green circles) have centers belonging to an ellipses (curved red line), the ellipse axes (straight red lines) are rotated  rotated by an angle $\delta$ respect to the vector ${\bf u}$ aligned with the $x$-axis.
({\bf{B}}) The Pearson correlation coefficient is computed as in Fig.~\ref{fig:Figure_2}  ($\sigma_{\lambda}/\lambda=0.01$, $\sigma_{\theta}=0.01$, $\sigma_{\epsilon}=0.06$).
  ({\bf{C}}) Correlation length for different values of $\sigma_{\epsilon}$ and $\sigma_{\delta}$ computed as in Fig.~\ref{fig:Figure_2} ($\sigma_{\lambda}/\lambda=0.01$, $\sigma_{\theta}=0.01$).
   ({\bf{D}}) Covering probability computed as in Fig.~\ref{fig:Figure_4} ($\sigma_{\lambda}/\lambda=0.01$, $\sigma_{\theta}=0.01$, $\sigma_{\epsilon}=0.06$).
We find that increasing variability reduces the correlation length  and decreases the covering probability.
Eq.~(\ref{eq:p_of_R}) correctly describes the  covering probability as a function of variance and environment size.
}
\label{fig:beta_epsilon}
\end{figure}

The triangular lattice defining the spatial activity of a grid cell is determined by linear combinations of two elementary vectors ${\bf{v}}$ and ${\bf{u}}$. 
The reference frame can be chosen to have the vector ${\bf{u}}$ coinciding with the 
$x$-axis, i.e.  ${\bf u}=\lambda_2(1, 0)$ and
the vector ${\bf v}=\lambda_1(\cos(\beta),\sin(\beta))$, where $\beta$ is the angle formed by the two elementary vectors (which can be restricted to the first quadrant). 
The two positive numbers $\lambda_2$ and $\lambda_1$ are the moduli of the two elementary vectors. 
The set of the six vertices defined by the triplet ${\bf u}$, ${\bf v}$, ${\bf u}-{\bf v}$ and their opposite vectors, forms an hexagon that can be inscribed into an ellipse centered at the origin, whose general equation is $Ax^2+2Bxy+Cy^2=1$ (see Fig.~\ref{fig:beta_epsilon}A). 
The (inverse squared) length of the two axes of the ellipse is determined by the eigenvalues of the quadratic form and their orthogonal directions are determined by the corresponding eigenvectors. 

An alternative parametrization of the ellipse is given by\,: 1) the direction $\delta$ of the axes of the ellipse with respect to the axes of the reference frame\,; 2) the ratio $\epsilon$  between the length of the two axes (i.e. the ellipticity of the grid as defined in~\cite{Stensola2012})\,; 3) the length  $\lambda/\sqrt{\epsilon}$ of the axis parallel to the $x$-axis when $\delta=0$ (the other axis has length $\sqrt{\epsilon}\, \lambda$).
By requiring that the ellipse pass through the three independent vertices ${\bf u}$, ${\bf v}$, ${\bf u}-{\bf v}$, we obtain the relations 
\begin{eqnarray}
&\lambda_2=\frac{\lambda}{\sqrt{F_1}}\,;\quad \lambda_1\sin\beta=\frac{\sqrt{3}}{2}\lambda_2F_1\,; \quad
\lambda_1\cos\beta=\frac{\lambda_2F_2}{2} \,;\nonumber \\
&F_1\equiv {\epsilon}\cos^2\delta+\frac{1}{\epsilon}\sin^2\delta\,;\nonumber \\
 &F_2\equiv 1-\sqrt{3}\left( \epsilon -\frac{1}{\epsilon}\right)\sin\delta\,\cos\delta \,,
\end{eqnarray}
which provide a general mapping between the free parameters $\lambda$, $\epsilon$, $\delta$ of the ellipse and the free parameters $\lambda_1$, $\lambda_2$, $\beta$ of the vectors ${\bf{u}}$ and ${\bf{v}}$. 
Ellipses with $\delta=0$ have axes aligned with the coordinate system.
Elementary algebra shows that this condition corresponds to isosceles triangles with $|{\bf v}|=|{\bf u}-{\bf v}|$, i.e. $2\lambda_1\cos\beta=\lambda_2$ or  $\cos\beta=1/{\sqrt{1+3\epsilon^2}}$.
The special case of $\epsilon=1$  fixes $\lambda_1=\lambda_2=\lambda$ and $\cos\beta=1/2$, i.e. corresponds  to equilateral triangles. 
Note that the direction $\delta$  is related only to the deformation of the hexagon defined by the elementary vectors and its variations do no affect the orientation  $\theta$ of the grid.

We generalize the analysis of the main text to cases where the axes of the ellipse are not aligned with the coordinate system ($\delta\not=0$), which generally corresponds to scalene triangles.
We choose a parametrization where  $\epsilon$, $\delta$  and $\lambda$ vary independently. 
The upshot is that the results presented in the main text still hold.
Specifically, for fixed $\sigma_{\lambda}/\lambda$ and $\sigma_{\theta}$  we compute the correlation coefficient between the number of neurons that are active at the center of two unit cells, as described in the main text. Fig.~\ref{fig:beta_epsilon} illustrates the results of the simulations for different values of $\sigma_{\epsilon}$ and $\sigma_{\delta}$ using general grids. As for the other sources of variability, the correlation decreases rapidly with the separation between the two centers and the correlation length decreases as $\sigma_{\epsilon}$ and $\sigma_{\delta}$ increase.
Finally, Fig.~\ref{fig:beta_epsilon}D also shows that the covering probability conforms to the relation (\ref{eq:p_of_R_of_L}) presented in the main text.

\section{Analysis of the correlation function with no orientation variance}\label{supp:correlation_function}
In the main text we showed that, as long as  there is some variability in the orientation, the correlation function  of the number of neurons active at two spatial points tends to zero as the distance increases. 
Here we discuss the asymptotic behavior of the correlation function in the case without orientation variance in the grid parameters. 

\begin{figure}[htb]
  \centering
   \includegraphics[keepaspectratio, keepaspectratio]{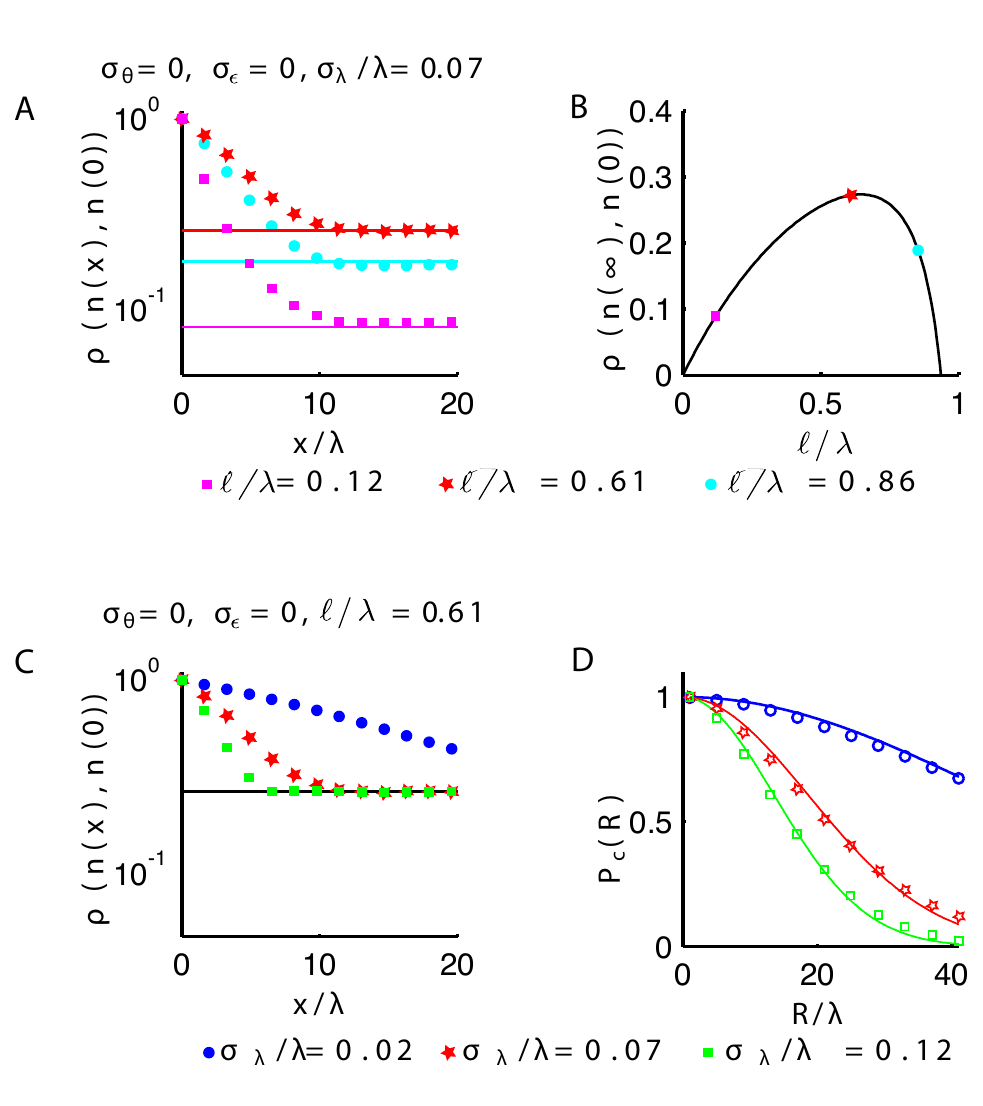}
\caption{\textbf{The correlation length depends on the behavior above the asymptotic value of the correlation.
}
(\textbf{A}) When there is no variance in grid orientations, the correlation function between the number of active cells at two locations, $\rho(n(x),n(0))$, approaches a nonzero asymptotic value that depends on $\ell/\lambda$.   The solid lines indicate the theoretical prediction of the asymptotic values from Eq.~(\ref{eq:supp_asymp}). Simulation parameters are $\sigma_{\lambda}/\lambda=0.07$, $\sigma_{\theta}=0$, $\sigma_{\epsilon}=0$. 
   (\textbf{B}) The asymptotic value of the correlation in the absence of orientation variance is  predicted by Eq.~(\ref{eq:supp_asymp}) (black line).   Representative values corresponding to the three curves in panel {\bf A} are marked by the colored symbles.   Note that there is a maximum value in the asymptotic correlation as a function of $l/\lambda$.
(\textbf{C}) The correlation decreases faster when the variance in the period increases ($\ell/\lambda=0.61$, $\sigma_{\theta}=0$, $\sigma_{\epsilon}=0$).
(\textbf{D}) Numerical simulations (colored symbols) determine the covering probability of the environment for the different variances in the grid parameters and environment sizes.   The numerics are accurately predicted by  Eq.~(\ref{eq:covering_probability}) of the main text (solid lines) in which the correlation length for a grid system was assessed as the distance at which $\rho$ in panel {\bf C} decreased to  $1/e$. This threshold is always larger than the asymptotic value of the correlation (see main text). 
}
\label{fig:corr_1}
\end{figure}

The derivation of the asymptotic correlation has been  outlined in Section~\ref{sec:dephasing}; the absence of orientation variance affects the computation of the conditional probability as follows.
We divide space into unit cells using the mean grid parameters and indicate with ${\boldsymbol{\phi}}_n$ the phase  in the unit cells centered at ${\bf y}=(n\lambda,0)$, $n=0, \, 1, \, \dots$.
In the case in which the grid has the same orientation as the mean grid but different period ($\lambda'\neq\lambda$), the phase ${\boldsymbol{\phi}}_n$ will gradually shift along the $x-$axis as $n$ increases but it will always belong to a one dimensional surface with fixed $y$ component  equal to ${\boldsymbol{\phi}}_0^y$.
For large $n$ the phase becomes randomly distributed along the segment $[(n-1/2)\lambda,(n+1/2)\lambda]$. 
The resulting probability that the point ${\bf y}$ is covered, given that ${\bf x}=(0,0)$ is covered, depends on ${\boldsymbol{\phi}}_0^y$. 
In particular, a grid that covers the point ${\bf x}$ will have an intersection  of length $2 \sqrt{\ell^2/4-\left(\phi_0^y\right)^2}$ between its firing field and the segment $[-1/2\lambda,+1/2\lambda]$.
Because of the previous argument, for large $n$ this interval will be uniformly distributed along the segment $[(n-1/2)\lambda,(n+1/2)\lambda]$ so it will cover the point ${\bf y}$ with probability 
\begin{equation}
Q_{\boldsymbol{\phi}}({\bf y}|{\bf{x}}) = \frac{  2 \sqrt{\ell^2/4-\left(\phi_0^y\right)^2}}{\lambda} \, .
\end{equation}
Furthermore, the probability to have a neuron active at ${\bf x}$ is
\begin{equation}
Q_{\boldsymbol{\phi}}({\bf{x}}) = 
  \begin{cases} 
   0 & \text{if }\left(\phi_0^x\right)^2+\left(\phi_0^y\right)^2>  (\ell/2)^2\,, \\
  1      & \left(\phi_0^x\right)^2+\left(\phi_0^y\right)^2 \leq  (\ell/2)^2 \,.
  \end{cases}
\end{equation}
Combining the previous results we obtain $\langle a({\bf{y}})a({\bf{x}})\rangle \to \frac{4 \sqrt{3} \ell^3}{9 \lambda^3}$.
Hence, if the orientation variance is zero the correlation coefficient between two distant points reaches  the asymptotic value
\begin{equation}\label{eq:supp_asymp}
\rho({a(\bf{\infty}}),a({\bf{0}}))=\frac{\frac{8 \ell}{3 \pi {\lambda}}-\frac{\pi \ell^2}{2 \sqrt{3} {\lambda}^2}
}{1-\frac{\pi \ell^2}{2 \sqrt{3} {\lambda}^2}} \, .
\end{equation}
This has been confirmed numerically in Fig.~\ref{fig:corr_1}.
The same argument holds if ellipticity variance is present.

In the main text we defined the correlation length of a grid system as the distance at which the correlation in the number of active cells falls below the threshold $1/e$.  
Results of the present Section qualify the range in which this threshold could be chosen.
Indeed, for a given  $\ell/\lambda$, Eq.~\eqref{eq:supp_asymp} gives the asymptotic value of the correlation function when no variance in the orientation is present, e.g. in the biological system $\ell/\lambda\approx1/1.63$ and the asymptotic value of the correlation  is about $ 0.28$ (see Fig.~\ref{fig:corr_1}B).
Because of the effect described above, if the threshold used to define the correlation length is chosen below this value, the correlation length will depend only on the orientation variance.  
Hence, in order to assess the length scale of correlation that is affected by the variance in all the grid parameters, a threshold slightly above the asymptotic value should be chosen (in the main text we used $1/e$.)
This choice is relevant because it captures the dependence of the covering probability on the variance in the grid parameters.
In fact, if our definition of the correlation length is used to analyze the covering of a system, the analytical results obtained from our approach are in agreement with direct numerical analysis performed with (see Fig.~\ref{fig:Figure_4}) and without  (see Fig.~\ref{fig:corr_1}D) orientation variance.

\section{Covering probability of a unit cell}\label{supp:p_1D}
We discuss the covering probability of a unit cell in one dimension. 
We take each grid cell to be active in intervals of length $\ell$, regularly spaced with centers at distance $\lambda$ apart.
A unit cell is given by an interval of length $\lambda$.
The covering probability of a unit cell by $N$ grid cells is analogous of that  of covering a region of length $\lambda$ by $N$ intervals of length $\ell$ with periodic boundary conditions on the region.
This probability distribution has been computed analytically in~\cite{Stevens} and reads
\begin{equation}\label{eq:Stevens}
P\left(N,\frac{\ell}{\lambda}\right)=\sum_{k=0}^{N }(-1)^{k}{{N}\choose{k}} f(k) \, ,
\end{equation}
where $ f(k)=\left(1-k{\ell}/{\lambda} \right)^{N-1}$ if $k{\ell}<{\lambda}$ and $ f(k)=0$ otherwise.
For large $N$ this relation reduces to the result used in the main text 
$P=1-N\left( 1-\ell/\lambda\right)^{N-1}$.
 
The proof of (\ref{eq:Stevens}) is presented below for the sake of completeness.  Because we have periodic boundary conditions on the region to be covered, we can regard it as a circle of unit length and we can take the intervals of length $\zeta=\ell/\lambda$ to be arcs on this circle.   The arcs are labelled by their order of occurrence in the anti-clockwise direction around the circle, starting by convention from the north pole. The arcs are identified by their initial position\,;  there is a gap after the $r$-th arc if the  distance between the initial positions of the $r$-th and the $r+1$-th arcs  is larger than the size of the arcs. For convenience, we rigidly translate all  the arcs so that the first one is positioned at the north pole -- this convention does not affect the probability of coverage. 

Consider $N$ random arcs of length $\zeta$ on the circle.  Draw $k$ arbitrary arcs from this set (say $(r_1,r_2,\ldots r_k)$, with $k\le N$). Let $f(k)$ be the probability that each arc in this randomly selected subset is followed by a gap, irrespective of the state (followed by a gap or not) of all the other arcs.  From the $f(k)$'s, the probability (\ref{eq:Stevens}) of leaving {\it no} gaps is computed as follows.  
First, let $Q(n_g,n_u)$ be the probability that  $n_g$ prescribed arcs are each followed by a gap and $n_u$ prescribed arcs are each not followed by a gap, with the rest of the $N$ arcs in unspecified states.    Then, $Q(n_g,1)=f(n_g)-f(n_g+1)$ because $f(n_g)$ includes the probability that the extra arc might be gapped or ungapped, while $f(n_g+1)$ subtracts the probability that the extra arc is in fact gapped.  By a similar reasoning we obtain $Q(n_g,2)=Q(n_g,1)-Q(n_g+1,1)$ and so on recursively up to $n_g+n_u=N$. Simple algebra shows then that the probability of leaving no gaps is
\begin{equation}
P\left(N,\frac{\ell}{\lambda}\right)=Q(0,N)=\sum_{k=0}^{N}(-1)^{k}{{N}\choose{k}} f(k)\,.
\label{eq:Steve}
\end{equation}

\begin{figure}[htb]
  \centering
    \includegraphics[keepaspectratio,keepaspectratio]{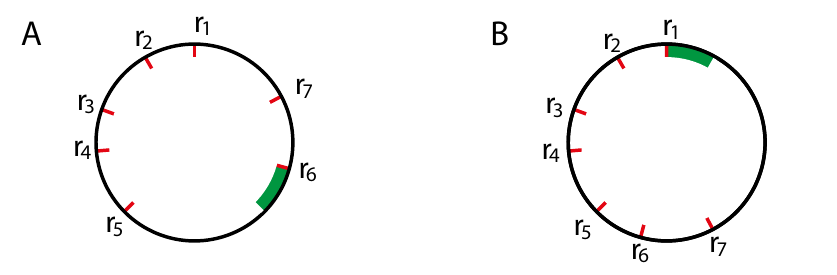}
  \caption{\textbf{Shift of the arcs following the gap affects only  the covering of the arc that is followed by the gap.}
 (\textbf{A}) Circle of unit length with seven arcs of which only the initial position has been represented (red lines  at angle $r_k$).
  Each arc has  length $\zeta=1/4$ and the 5-th arc is followed by a gap (green region).
  (\textbf{B})  the 6-th and 7-th arc have been rotated clockwise by an angle $\pi/2$ that corresponds to an arc of length $\zeta$.
  The rotation shifts the gap to the region before the first arc and does not affect the state of the 6-th and 7-th arc.}
  \label{fig:circle}
\end{figure}

The formula (\ref{eq:Steve}) leaves us to determine the expression of $f(k)$, which is done as follows. When an arc, say $r$, is followed by a gap, we rigidly shift backward (clockwise) all the following arcs up to the last ($N$-th) by an amount $\zeta$. Because the $r$-th arc is followed by a gap, the state of all the arcs other than $r$ is not affected by this backward shift and we are left with a final region of size $\zeta$ that does not contain any initial position of the arcs (see Fig.~\ref{fig:circle}). Note that whether the $N$-th arc is gapped or ungapped before this shift 
corresponds to whether or not the last arc partially overlaps with the final region of size $\zeta$. 
The probability of distributing $N-1$ initial positions of the arcs 
(other than the first one fixed at the origin) in a region of size $1-\zeta$ gives $f(1)=\left(1-\zeta\right)^{N-1}$. The reasoning for
$f(2)$ is similar. If the two prescribed gapped arcs are $r_1$ and $r_2>r_1$, we first shift backward by $\zeta$ all the arcs 
following $r_1$ and then again by $\zeta$ those following $r_2$. We are then left with a final
unoccupied region of size $2\zeta$. The crucial point is that the state of all the arcs other than $r_1$ and $r_2$ is again unaffected. We can then compute $f(2)=\left(1-2\zeta\right)^{N-1}$ as the probability of distributing $N-1$ initial positions of the arcs in the available region of size $1-2\zeta$. 
Generalizing the reasoning to $k$ arcs gives the expression $f(k)=\left(1-k\zeta\right)^{N-1}$, provided the total length of the arcs is smaller than the length of the circumference, i.e. $k\zeta\le 1$, otherwise $f(k)=0$. That completes the proof and yields Eq.~\eqref{eq:Stevens}.

\bibliographystyle{apsrev4-1} 
\bibliography{mybib} 

\end{document}